\documentclass[10pt]{article}
\usepackage{graphicx}
\usepackage{setspace}
\begin{document}

\begin{center}

\begin{Large}

\noindent
{\bf Reply to Comment on ``Drip Paintings and Fractal Analysis'' by Micolich 
{\em et al.} }
 
\end{Large}

\vspace{2mm}

\begin{large}

{\bf 
(arXiv:0712.165v1 [cond-mat.stat-mech])}

\vspace{3mm}

Katherine Jones-Smith, Harsh Mathur and Lawrence M. Krauss

\end{large}

\vspace{2mm}

{\em Department of Physics, Case Western Reserve University, 10900 Euclid 
Avenue, Cleveland, Ohio 44106-7079, USA}.

\end{center}

\vspace{4mm}

Recently we demonstrated that the fractal analytic techniques of Taylor {\em et al.} provide no information about whether a painting is an authentic Pollock \cite{usprep}.
Micolich {\em et al.} dispute our findings in a comment \cite{micolich},  and it is this comment which we now address. 

In a scientific exchange typically
all parties have a common interest in testing a hypothesis via experiments
with reproducible results. Unfortunately, this does not characterize the Micolich {\em et al.} comment, nor the exchanges that  preceded it.
As a common standard, data are made public and methods
are presented transparently. After nine years and numerous
publications Taylor {\em et al.}
have not provided even a simple tabulation of the paintings they have
analyzed and the basic fractal parameters for each. There is no adequate
description of their color separation methods, nor any independent 
corroboration of their results\footnote{Mureika {\em et al.} 
\cite{mureika} have also carried out box-counting analysis
of Pollock paintings but report that due to lack of sufficient resolution 
they do not find two fractal dimensions  
in any of the paintings they have analyzed. Observation
of two dimensions is the essential first step in the analysis
of Taylor {\em et al.} Thus the work of Mureika {\em et al.} does
not corroborate the findings of Taylor {\em et al.}}.
Even the number of
paintings they claim to have analyzed decreases occasionally, without 
justification\footnote{In \cite{taylorsciam} 
Taylor reports they have analyzed 20 known Pollock paintings. Four years later,
in a preprint version of  \cite{taylorpattern} 
they say they are building on their previous work in which they analyzed 5 known Pollocks, and therein they will present results for 17 newly analyzed Pollocks. However in the published version of 
\cite{taylorpattern}, they state they have analyzed only 14 paintings.}.
They have ignored requests by ourselves and other researchers to share minimal information (such as the basic parameters of the paintings they analyzed). Yet in print they profess an interest in encouraging further research in this area. 

Our e-print \cite{usprep} provides a detailed tabulation of our data, and a comprehensive description of our methods, with the aim of providing other researchers 
the information they would need to reproduce and verify our results. 
It is ironic that Micolich {\em et al.} \cite{micolich} criticize our work (incorrectly, 
as we show below) on the basis of this full access to our data, 
access they have not themselves granted to the community.

In \cite{usnature}, 
we identified a number of severe logical inconsistencies in the underlying theoretical framework  from which the  original hypothesis of ``fractal expressionism'' was set forth.
In our recent e-print \cite{usprep} we have shown
the unfortunate results that ensue should one choose to overlook these inconsistencies and engage fractal analysis as an authentication tool.  The onus now lies with the proponents of fractal expressionism, first, to present their data and methods, and, second, to squarely address the criticisms we presented.  

Moving on to the specific points raised by Micolich {\em et al.} in their recent
post, we begin by addressing the issue of whether we have overstated their
claims and used fractal analysis as a stand-alone 
``black-box authenticator''. Rather than
wrangle over whether we have exaggerated their claims, 
Taylor {\em et al.} should determine which of their claims they are willing
to stand by in light of our results, and then supply reproducible empirical arguments in support of them.
Our data make it clear that the fractal criteria of Taylor {\em et al.}
should play {\em no} role whatsoever in authenticity debates. Given the 
complete lack of correlation between artist and fractal characteristics that we
have found\footnote{Of the 3 authentic Pollocks we analyzed, 2 failed to be authentic according to the fractal criteria. Both of the amateur paintings we analyzed, created in 2007 by local artists, passed as authentic Pollocks according to the fractal criteria. Of the two paintings from the Matter cache, one passed and one failed. Additionally, many crude sketches created by one of us (KJS) 
pass the criteria, although they do not even resemble Pollock's work.
See \cite{usprep} and \cite{usnature} for details.}, 
in particular, the failure of 
fractal analysis to detect deliberate forgery, it is clear that box-counting
data are not useful even as a supplement to other analysis.

Debating whether their claims are more modest than our
paraphrasing misses the point. Nonetheless, we feel 
compelled to point out that the past claims of Taylor {\em et al.} regarding the use of fractal analysis as an authentication tool have not 
been particularly modest.  For example, regarding the monetary importance of distinguishing authentic Pollocks from fakes, Taylor writes in the introduction of 
\cite{santafe} that
{\em ``When dealing with such staggering commercial considerations,  subjective judgements attempting to identify the `hand' of the artist
may no longer be adequate. I will therefore demonstrate the considerable
potential that the fractal analysis technique has for detecting the `hand'
of Pollock by examining a drip painting which was sent to me to establish its authenticity.''} 
Presumably the inadequate ``subjective judgements'' Taylor refers to are 
provenance and connoisseurship, the same authentication tools that they
espouse with such zeal in their comment.
In the conclusion of the same paper Taylor asserts
{\em ``Therefore, fractality can be identified as the `hand' of Pollock and a fractal analysis can be used to authenticate a drip painting.''}
Similarly in ref \cite{taylorsciam} he writes
{\em ``We could therefore conclude that each of the five paintings sent
to us for analysis was produced by someone other than Pollock.
Fractality, then, offers a promising test for authenticating
a Pollock drip painting."}   Many other quotes to this effect can be found in their papers. Even the more modest position on authentication which Micolich {\em et al.} 
seem to adopt in their comment must address the shortcomings of fractal analysis we have identified\footnote{See previous footnote.}. 

Another point raised by Micolich {\em et al.} concerns overlapping
fractals. In \cite{usnature} we showed that when two ideal fractals (e.g.
Cantor dusts) are superimposed, the composite and the visible part of 
the lower layer are not fractals. In their comment, Micolich {\em et al.}
claim to have previously rebutted our results. In the same sense as we argued that
their commentary does not rise to the level of a scientific debate, we do
not feel that their response to this issue rose to the level of a rebuttal;
we present it here in full so that readers might quickly judge for
themselves. Regarding our proposition that it is impossible for 
superposed ideal fractals to yield fractal composites, they write:
{\em ``It is both mathematically and physically possible for two 
exposed patterns and their composite all to be fractal. This depends
on the relative densities of the exposed patterns and
the scaling behaviour of boxes containing both patterns. Jones-Smith
and Mathur's Cantor dust does not apply to Pollock paintings, where 
the overlap of layers is considerably more complicated''} \cite{themnature}. 
Next they
claim that in our analysis of the blue, black and blue-black composite
layers of {\em The Wooden Horse: Number 10A, 1948}, we ``find all three layers to be fractal'', 
thereby contradicting our earlier work on fractal superpositions.
But we have found no such thing. A key point of our earlier paper \cite{usnature}
was that analysis of a box-counting curve over less than two 
orders of magnitude, in the absence of any theoretical prediction
of scaling behavior, is an insufficient criterion by which to 
establish fractality. Fractal dimensions determined over such a 
limited range are meaningless. Thus we do not find the 
colored layers of {\em Wooden Horse} to ``be fractal'' any more than we find the
entire content of childish sketches such as {\em Untitled 5} to ``be fractal''. 
As stated in our earlier paper, the claim by Taylor {\em et al.}, that 
different colored layers appear to have different fractal dimensions,
is an artifact of the limited range over which Pollock paintings
permit box-counting analysis. We draw attention to the fact that the
range spanned by the largest of Pollock's paintings (3 orders of 
magnitude) is used to determine {\em two} independent fractal 
dimensions, thus each dimension is determined over $< 2$ orders
of magnitude, a range deemed insufficient by {\em both} sides of
a prominent debate \cite{debate} regarding the appropriate 
range needed to establish fractality. 

Micolich {\em et al.} declare that it is impossible for
a properly written boxcounter to produce a fractal dimension
$D > 2$ for a planar fractal. Since we obtain some dimensions
slightly greater than 2, they claim this proves our boxcounter
is flawed. It should be evident to anyone familiar with elementary data
analysis that their assertion is wrong; nonetheless, we now demonstrate
in detail that it is in fact entirely
possible to obtain $D > 2$.

\begin{figure}
\begin{center}
\includegraphics[width = 0.9\textwidth]{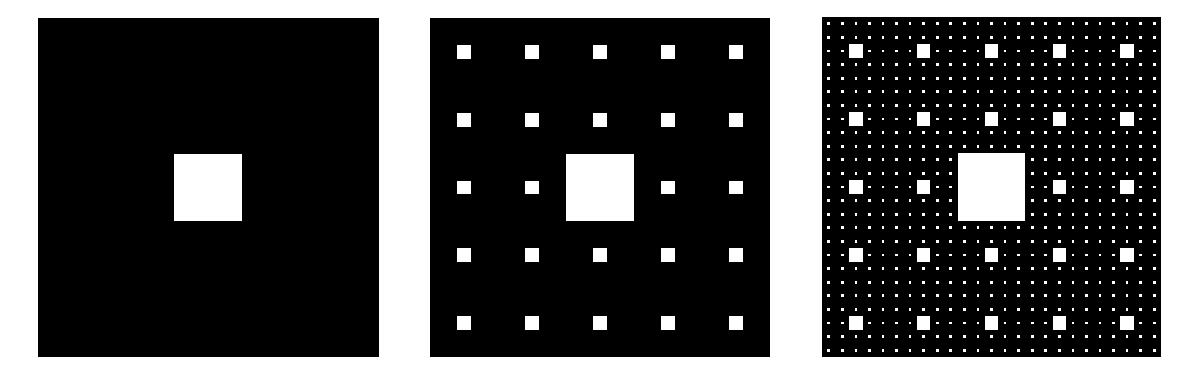}
\end{center}
\caption{{\bf The first three iterations of a base-5 Sierpinski carpet.} In the first 
iteration a unit square is subdivided into 25 sub-squares and the middle square is
deleted, in the second iteration this process is repeated with the 24 squares that
are retained, and so on. The exact fractal dimension of the base-5 Sierpinski
carpet is $1.974\ldots$}
\label{fig:carpet5}
\end{figure}

In a box-counting calculation one covers the canvas with
boxes of size $l$ and counts the number of filled boxes $N$.
For a fractal, the smoothed variation of $N$
with $l$ is a power law; the exponent is the boxcounting
dimension, $D$. Any real measurement involves uncertainty; determination
of $D$ is no exception. The sources of error
in this case are: (1) the boxcounts are discrete and deviate from a
smooth fit through the boxcounting curve ($N$ vs $l$ plot); (2) 
roundoff errors associated
with digital blurring of the image; and, (3) offset errors associated
with a mismatch between the size of the boxes and the 
canvas\footnote{Our
boxcounter uses the roundoff rule that a pixel occupies a box if more than 
half the area of the pixel lies inside the box. It places the origin of the grid
of boxes at the lower left corner of the canvas; boxes on the top and right
edge that lie partly outside the canvas are not counted. We have
explored other common variations such as including pixels in a box if they intersect
it at all; treating the edge boxes on the same footing as the interior boxes;
and offset averaging, where we shift the origin of our box-grid and average
over different locations. We have verified that all these alternatives lead to
essentially the same results.}.
Thus it is reasonable to expect that (a) the measured $D$ would deviate slightly
from the true ideal value; (b) as the range of box sizes measured is increased,
the deviation from the true value should
decrease; and (c) if the true fractal dimension is close to 2, we may
obtain a dimension slightly greater than 2. 

For illustration we present two examples. Table \ref{tab:dmoretwo} shows that for the
Sierpinski carpet ($D_{{\rm exact}} = \ln 8/\ln3 = 1.89...$)
at low resolutions we obtain $D > 2$; but as the resolution and range of
box sizes increases, the measured $D$ converges steadily to $D_{{\rm exact}}$.
The Sierpinski carpet is constructed by subdividing a square into $3^2$ 
squares, removing the central square, and reiterating. We have
considered two variations based on fifths ($D_{{\rm exact}} = 1.974...$;
see fig \ref{fig:carpet5})
and sevenths ($D_{{\rm exact}} = 1.989...$) instead of thirds. In these
cases $D > 2$ is obtained even at high resolution, and over a much 
bigger range of box sizes than in the conventional Sierpinski carpet,
which has a smaller $D_{{\rm exact}}$.
An even simpler example is a filled canvas ($D_{{\rm exact}} = 2$). 
In this case it is not necessary to use a boxcounting program; the box-counts
are given by the formula $N = [{\cal F}(L/l)]^2$ where $L$ is the size of the canvas
and ${\cal F}(x)$ is the biggest integer less than or equal to $x$. In this case
too we obtain $D > 2$ (see Table \ref{tab:dmoretwo}) 
and since the results are based on an 
analytic formula, they are independent of any possible errors in our 
boxcounter\footnote{The convergence to $D_{{\rm exact}}$ for 
the Sierpinski carpet shows that our boxcounter is working. 
We have tested it rigorously. Among many
other checks, we have
verified that it gives expected results for a point, a line, a filled canvas, 
four variations on the Sierpinski carpet (each with a distinct dimension),
and a Koch snowflake. We have also checked the box-counts directly.
For a Sierpinski carpet, the counts can be calculated analytically, if the
box-size is commensurate with the carpet; for a filled canvas and line
they may be calculated even in the incommensurate case. For low resolution
carpets with large boxes the counts can be worked out by hand even
in the incommensurate case. Our boxcounter provides exactly the right
count in all these cases.}.

\begin{table}
\begin{center}
\begin{tabular}{|c|c|c|c|c|}
\hline
Image size & $l_{{\rm min}}$ & $l_{{\rm max}}$ & $D_{{\rm sierpinski}}$ &
$D_{{\rm filled}} $ \\
(pixels) & (pixels) & (pixels) & & \\
\hline
$243 \times 243$ & 10 & 22 & 2.07 & 2.14 \\
$729 \times 729$ & 10 & 86 & 1.97 & 2.06 \\
$2187 \times 2187$ & 10 & 213 & 1.92 & 2.02 \\
$6561 \times 6561$ & 10 & 591 & 1.92 & 2.02 \\
\hline
\end{tabular}
\end{center}
\caption{Fractal dimension for Sierpinski carpet and completely filled
canvas images of different resolutions. $D_{{\rm exact}} = 1.89\ldots$
for the carpet and $2$ for the filled canvas. The measured dimension $D > 2$ 
for the carpet at low resolution; for the filled canvas, $D > 2$ always. $l_{{\rm min}}$
is the smallest box size considered, $l_{{\rm max}}$ is the largest.}
\label{tab:dmoretwo}
\end{table} 

Finally, we note that a measured $D > 2$ does not imply 
boxes are being overcounted. We have verified in
cases where $D > 2$
that the number of filled boxes returned by our boxcounter
never exceeds the total number of boxes. For the filled canvas
the number of filled boxes exactly equals the total number of boxes;
yet the measured $D > 2$.
Indeed by itself, the constraint
that the filled boxes must be fewer in number than the
total number of boxes, does not require that the local
slope of the boxcounting curve has to be less than the 
two.


\begin{thebibliography}{99}

\bibitem{usprep} Katherine Jones-Smith, Harsh Mathur and Lawrence M. Krauss,
{\em Drip Paintings and Fractal Analysis}, arXiv:0710.4917v2 [cond-mat.stat-mech].

\bibitem{micolich} A.P. Micolich, B.C. Scannell, M.S. Fairbanks, T.P. Martin,
and R.P. Taylor, arXiv:0712.1652v1 [cond-mat.stat-mech].

\bibitem{mureika} J.R. Mureika, C.G. Dyer and G.C. Cupchik, 
Phys Rev {\bf E72}, 046101 (2005).

\bibitem{taylorsciam} Richard P. Taylor, {\em Order in Pollock's Chaos}, 
Scientific American, Dec 2002.

\bibitem{taylorpattern} R.P. Taylor et al., {\em Authenticating Pollock Paintings Using Fractal Geometry}, Pattern Recognition Letters 28, 695 (2007)

\bibitem{usnature} Katherine Jones-Smith and Harsh Mathur, {\em Fractal Analysis: Revisiting Pollock's drip paintings}, Nature {\bf 444}, 
doi:10.1038/nature05398 (2006).

\bibitem{santafe} R.P. Taylor, ``Fractal 
Expressionism---Where Art Meets Science'', p 117,
in {\em Art and Complexity}, J. Casti and A. Karlqvist (eds) (Elsevier
Press, Amsterdam, 2003).

\bibitem{themnature} R.P. Taylor, A.P. Micolich and D. Jonas, {\em
Fractal Analysis: Revisiting Pollock's drip paintings (Reply)}, Nature
{\bf 444}, doi:10.1038/nature05399

\bibitem{debate} Avnir, D., Biham, O., Lidar, D. and Malcai, O., {\em Is the Geometry of
Nature Fractal?}, Science {\bf 279}, 39 (1998); Mandelbrot, B.B., {\em Is Nature
Fractal?}, Science {\bf 279}, 783 (1998).

\end{thebibliography}
\end{document}